# AN EXTENSIVE ANALYSIS OF QUERY BY SINGING/HUMMING SYSTEM THROUGH QUERY PROPORTION

Trisiladevi C. Nagavi[1] and Nagappa U. Bhajantri[2]

[1]Department of Computer Science and Engineering, S.J.College of Engineering, Mysore, Karnataka, India
tnagavi@yahoo.com

[2]Department of Computer Science and Engineering,
Government Engineering, College Chamarajanagar, Karnataka, India
bhajan3nu@gmail.com

## ABSTRACT

*Query by Singing/Humming (QBSH) is a Music Information Retrieval (MIR) system with small audio excerpt as query. The rising availability of digital music stipulates effective music retrieval methods. Further, MIR systems support content based searching for music and requires no musical acquaintance. Current work on QBSH focuses mainly on melody features such as pitch, rhythm, note etc., size of databases, response time, score matching and search algorithms. Even though a variety of QBSH techniques are proposed, there is a dearth of work to analyze QBSH through query excerption. Here, we present an analysis that works on QBSH through query excerpt. To substantiate a series of experiments are conducted with the help of Mel-Frequency Cepstral Coefficients (MFCC), Linear Predictive Coefficients (LPC) and Linear Predictive Cepstral Coefficients (LPCC) to portray the robustness of the knowledge representation. Proposed experiments attempt to reveal that retrieval performance as well as precision diminishes in the snail phase with the growing database size.*

## 1. INTRODUCTION

The MIR research focuses on use cases, nature of query, the perception of match, and the form of the output. Queries and output can be in textual form such as meta-data, music portions, soundtracks, scores, or music features. The matching process may retrieve music with specific content, or retrieve near neighbors from music database [1]. Most of the research works on QBSH [2, 3, 4, 5] are based on music processing and focused on many components like melody extraction, representation, similarity measurement, size of databases, query and search algorithms.

Here, we are proposing an analysis on QBSH through query excerption. To start with devotional music database is created; descriptive features like MFCC, LPC and LPCC are extracted and aggregated in symbolic strings representation. Later these strings are used for pattern matching using similarity measures. For audio retrieval, there are three similarity measures employed such as Euclidean Distance (ED), K- Nearest Neighbor (k-NN) and Dynamic Time Warping (DTW).

However, music usually consists of instruments and voices playing harmonically or in opposition to each other at the same time. In real audio recordings, audio information of all instruments and voices are mixed and stored in all channels. However, user of a QBSH system desires to query the music sung by the lead voice or played by a solo instrument. Therefore, the recordings need to be reduced to a more accurate representation of components related to music.

Proposed method has attempted to represent the QBSH analysis system using different feature extraction algorithms and similarity measures. Before the introduction of the methodology, in section 2 related work is discussed. Section 3 and 4 describe the experiments conducted to evaluate our approach. The results and discussions of the experiments are presented in section 5. Subsequently paper is concluded in section 6.

## 2. RELATED WORK

In modern days traditional ways of listening to music, and methods for discovering music, are being replaced by personalized ways to hear and learn about music. In other words, it is changing the nature of music dissemination [1]. The rise of audio and video databases necessitate new information retrieval methods tailored to the specific characteristics and needs of these data types. An effective and natural way of querying a musical audio database is by singing/humming the tune of a song. Even though a variety of QBSH [2, 3, 4, 5] techniques have been explored, there has been relatively less work on analysis of QBSH system through query excerption.

Many QBSH techniques represent song or piece of song as point sequences [6], Hidden Markov Models (HMMs) [7], Modified Discrete Cosine Transform (MDCT) coefficients and peak energy [8]. Few [9] use loudness model of human hearing perception along with local minimum function to recognize onsets in music database and query. In other works[10] music is treated as a time series and employed a time series matching approach because of its effectiveness for QBSH in terms of robustness against note errors.

In [11], authors retrieve the melody of the lead vocals from music databases, using information about the spatial arrangement of voices and instruments in the stereo mix. The retrieved time series signals are approximated into symbolic strings which reveal higher-level context patterns. The method elaborated in [12] builds an index of music segments by finding pitch vectors from a database of music segments. In another work [2], a method for showing the melodic information in a song as relative pitch changes is projected. Further, the work [13] is based on extracting the pitch out of monophonic singing or humming, and later segmenting and quantizing the information into a melody composed of discrete notes.

Some [4, 6, 12] of them concentrate on developing QBSH systems using mammoth music collections. The objective is to construct a dependable and well-organized large-scale system that collects thousands of melodies and responds in seconds. The query melody searching and alignment is done using skeletons of the melody [6]. To compliment the work [14] proposed a technique using melody matching model based on the genetic algorithm and improving the ranking result by Local Sensitive Hashing (LSH) algorithm. Authors [8] employed double dynamic programming algorithm for feature similarity matching.

The melody similarity measure [6] is then derived based on the arrangement of the point sequences. Authors claim their method is robust against pitch errors and tempo variations in the queries, which is especially advantageous for QBSH. In [7] query is judged similar to the database melody if it's HMM has a high likelihood of generating the query. Authors [10] have explored the responsiveness of note-interval dynamic programming searches to different parameters and portrayed two-stage search merging fast n-gram with a more precise but slower dynamic programming algorithm. Consequently to become accustomed to people's

singing\humming practice, a new melody representation and new hierarchical matching methods are proposed [12].

In retrieval, the system without human intervention transliterates a sung query into notes and then digs out pitch vectors similar to the index construction. For each query pitch vector, the technique searches for similar melodic segments in the database to obtain a list of contender melodies. This is performed efficiently by using LSH [14]. Melody is compared to a database of indexed melodies using an error tolerant similarity search [13]. Further authors [13] proposed two algorithms for melody extraction, roughly characterized by the trade-off between accuracy of transcription also called as recall and computing time needed. In [11] system represents musical tunes as time series and uses time warping distance metric for similarity comparisons. To substantiate a multidimensional index structure is used to prune the search space of songs and efficiently return the top hits back.

Some methods have small retrieval precision as they rely on melodic contour information from the song\hum tune, which in turn relies on the error-prone note segmentation procedure. Several systems yield improved precision when matching the melody directly from music, but they are slow because of their widespread use of DTW. Previous QBSH systems have not concentrated on analysis through query excerption for various reasons. Our approach attempts to portray the importance of both the retrieval precision and query excerption in QBSH systems.

## 3. METHODOLOGY

The design of QBSH system consists of two distinguished phases. The first phase is training, while the second one is referred to as operation or testing phase as described in figure 1. Each of these phases perform different operations on the input signal such as Pre-processing, Vocal and Non-Vocal Separation, Feature Extraction and Query Matching. The QBSH system steps are discussed below.

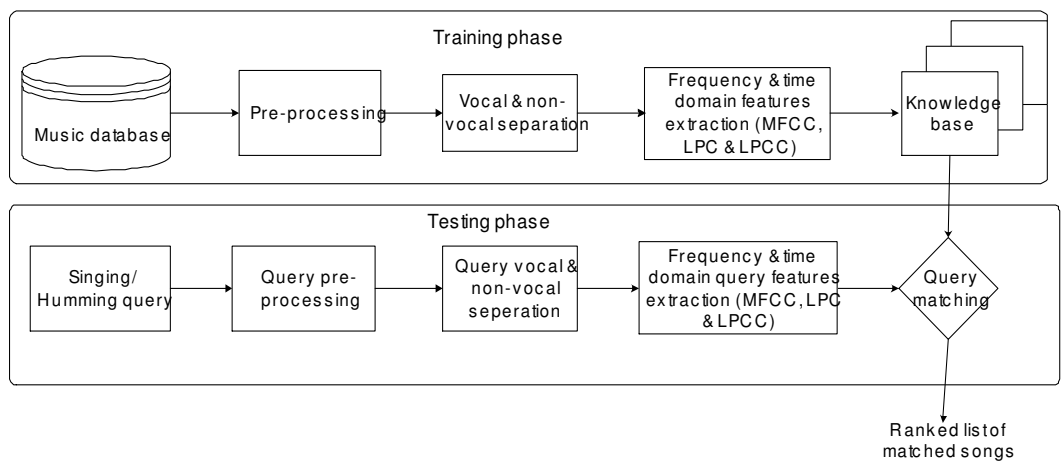

Figure 1. Block Diagram of Proposed QBSH System

## 3.1. Pre-processing

MP3 songs contain convoluted melody information and even noise. Thus pre-processing is applied on the MP3 songs database to extract information needed by the system. Most of the MP3 songs possess 44.1 KHz sample rate and dual-channel data, but for melody representation, such high quality signal is not necessary. Also it will make further processing time consuming and inefficient. In fact, even in very low sample rate, melody of the songs can be identified.

Therefore, MP3 songs are decoded into wave streams, down sampled to 8 KHz and converted to mono channel.

## 3.2. Vocal and Non Vocal Separation

In music, human vocal part always plays an important role in representing melody rather than its background music, it is desired to segregate both. Furthermore, music researchers have shown that the vocal and non-vocal separation exploits the spatial arrangement of instruments and voices in the stereo mix and could be described as inverse karaoke effect. Most karaoke machines adopt centre pan removal technique to remove the lead voice from a song [11]. One stereo channel is inverted and mixed with the other one into a mono signal. The lead voice and solo instruments are generally centered in the stereo mix whereas the majority instruments and backing vocals are out of centre. Above mentioned transformation is used to remove the lead voice from music. We intend to invert this effect, so that the pre-processed song yields a high portion of the lead voice, while most other instruments are isolated and removed. For vocal and non-vocal separation audio editor's voice extractor option with centre filtering technique is employed.

## 3.3. Feature Selection and Extraction

Extracting significant feature vectors from an audio signal is a major task to produce a better retrieval performance. In this work, Mel Frequency Cepstral Coefficients (MFCC), Linear Predictive Coefficients (LPC) and Linear Predictive Cepstral Coefficients (LPCC) features are preferred because they are promising in terms of discrimination and robustness. For these features frame size and hop size that is the span between the starting times of two succeeding frames are empirically determined for the retrieval. Most feature extraction techniques produce a multidimensional feature vector for every frame of audio.

### 3.3.1. Mel-Frequency Cepstral Coefficients (MFCC)

MFCC is based on the information conceded by low-frequency components of the audio signal, in which less emphasis is placed on the high frequency components. The aim of MFCC is to produce best approximation of the human auditory system's response. Further, MFCC is based upon short-time spectral analysis in which MFCC vectors are computed. The overall process of the MFCC is shown in figure 2.

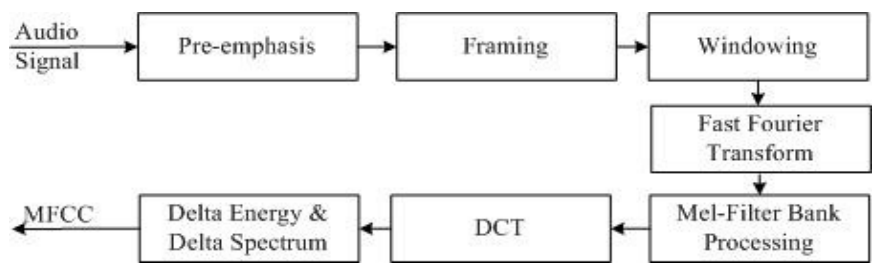

Figure 1: MFCC Block Diagram.

As shown in figure 2, MFCC consists of seven computational steps. Each step has its significance and mathematical approach as discussed below:

**Step 1:** Pre-emphasis refers to a procedure intended to amplify the energy of signal at higher frequencies with regard to the energy at lower frequencies in order to improve the overall signal to noise ratio in subsequent parts of the system.

**Step 2:** In framing, the pre-emphasized signal is split into several frames, such that each frame is analysed in short time instead of analyzing the entire signal at once. Usually the frame length is in the range 10 to 30 msec at which most part of the audio signal is stationary. Also an overlapping is applied to frames due to windowing in which will get rid of some of the information at the beginning and end of each frame. Overlapping reincorporates the information back to extracted features.

**Step 3:** The purpose of applying Hamming window is to minimize the spectral distortion and the signal discontinuities. Windowing is a point wise multiplication between the framed signal and the window function. Further the crisp representation of the Hamming window procedure is as follows:

$$W(n) = 0.54 - 0.46\, cos\left[\frac{2\pi n}{N-1}\right], where\ 0 \le n \le N-1 \quad (1)$$

If the window is defined as $W(n),\ 0 < n < N-1$

where N, Y[n], X[n], W[n] are number of samples in each frame, output signal, input signal and Hamming window respectively.

Then the result of windowing signal is as follows:

$$Y(n) = X(n) \times W(n) \qquad\qquad (2)$$

**Step 4:** The purpose of FFT is to convert the signal from time domain to frequency domain preparing to the next stage that is Mel frequency wrapping. The basis of performing Fourier transform is to convert the convolution of the glottal pulse and the vocal tract impulse response in the time domain into multiplication in the frequency domain. The equation is given by:

$$Y(w) = FFT[h(t) \star X(t)] = H(w) \star X(w) \qquad (3)$$

If X (w), H (w) and Y (w) are the Fourier Transform of X (t), H (t) and Y (t) respectively.

**Step 5:** Audio signal's low frequency components are more important than high frequency components. In order to highlight the low frequency components, Mel scaling is performed. Mel filter banks are non-uniformly spaced on the frequency axis, so we have more filters in the low frequency regions and less number of filters in high frequency regions. After having the spectrum from FFT for the windowed signal, Mel filter banks are applied, which resembles the human ear response. Filter banks can be implemented in both time domain and frequency domain. For the purpose of MFCC processing, filter banks are implemented in frequency domain. The filter bank has a triangular band pass frequency response and spacing. Further the bandwidth is determined by a constant Mel-frequency interval as shown in figure 3.

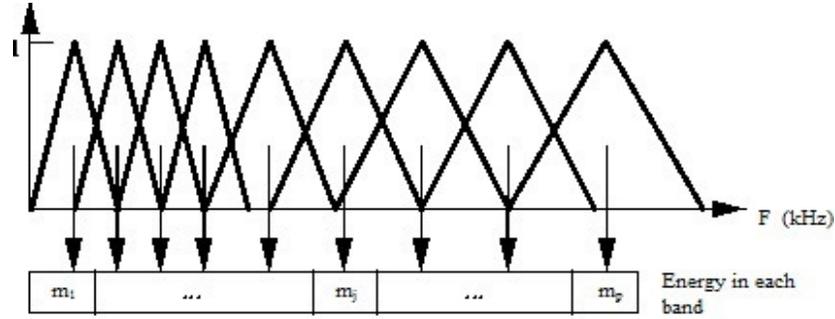

Figure 2: Mel Scale Filter Bank.

After that the approximation of Mel from frequency can be expressed as:

$$F(Mel) = [2595 \star log10 [1 + f] 700] \qquad (4)$$

where f denotes the real frequency and F (Mel) becomes the perceived frequency.

**Step 6:** This step performs conversion of the log Mel spectrum into time domain using DCT. The outcome of the conversion is called Mel Frequency Cepstrum Coefficient. The set of coefficients is referred as acoustic vector. Hence, each audio input is transformed into a series of acoustic vector.

**Step 7:** The audio signal and the frames vary such as the slope of a formant at its transitions. Consequently, this necessitates the addition of features correlated to the change in cepstral features over time. Thus 13 delta or velocity features that is 12 cepstral features plus energy and 39 double delta or acceleration feature are added. The energy in a frame for a signal x in a window from time sample t1 to time sample t2, is represented as following:

$$Energy = \sum X^2 [t] \qquad (5)$$

All 13 delta characteristics signify the transformation between frames referred as cepstral or energy features as stated in equation (6), while each of the 39 double delta characteristics signify the variation between frames in the resultant delta features.

$$d(t) = \frac{c(t+1) - c(t-1)}{2} \qquad (6)$$

### 3.3.2. Linear Predictive Co-efficients (LPC)

LPC analysis is to represent each sample of the signal in the time domain by a linear combination of p preceding values s (n-p-1) through s (n-1). Here p is the order of the LPC analysis. In the proposed approach, LPC analysis uses the autocorrelation method [15] of order p with value 14. The LPCC extraction procedure is portrayed in figure 4. Pre-emphasis, Framing and Windowing steps are identical to the algorithm discussed in section 3.3.1. Procedure contemplates directly with Linear Prediction Autocorrelation Analysis.

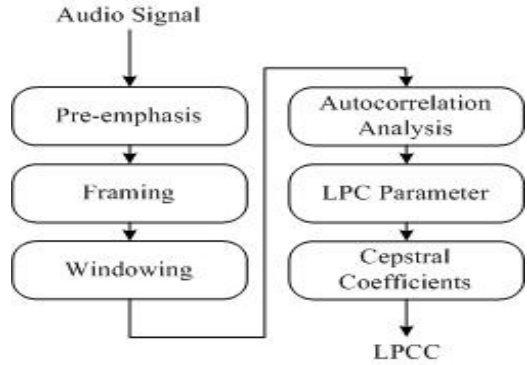

Figure 3: Linear Prediction Analysis Block Diagram.

The frame x (n) is assumed to be zero for n<0 and n>=N by multiplying it with Hamming window, the error reduction of pth-order linear predictor produces the well-known normal equations. Equations (7) and (8) are shown as below:

$$\sum_{k=1}^{p} a_k R(\mid i-k \mid) = R(i), \ 1 \le i \le p \qquad (7)$$

where

$$R(i) = \frac{1}{L-1} \sum_{n=0}^{N-1-i} y_n y_{n+i} \qquad (8)$$

The coefficients R (i-k) form an autocorrelation matrix which is a symmetric Toeplitz matrix. Toeplitz matrix is a matrix wherein all the elements along each descending diagonal from left to right are equal. Using matrix form considerably simplifies the combination (7) and (8) into

$$Ra = r \qquad (9)$$

where

$$r = [r(1)r(2)\ldots r(p)]^T \qquad (10)$$

r is a p×1 autocorrelation vector,

$$a = [a_1 a_2 \ldots a_p]^T \qquad (11)$$

a is a p×1 predictor coefficients vector and

$$\begin{pmatrix} r(0) & r(1) & r(2) & \ldots & r(p-1) \\ r(1) & r(0) & r(1) & \ldots & r(p-2) \\ r(2) & r(1) & r(0) & \ldots & r(p-3) \\ \ldots & \ldots & \ldots & \ldots & \ldots \\ r(p-1) & r(p-2) & r(p-3) & \ldots & r(0) \end{pmatrix} \qquad (12)$$

r is the p×p Toeplitz autocorrelation matrix, which is non-singular. In order to find a predictor coefficient vector, we need to solve linear system by a matrix inversion.

$$a = R^{-1}r \qquad (13)$$

### 3.3.3. Linear Prediction Cepstral Coefficients (LPCC)

LPCCs are the coefficients of the Fourier transform representation of the logarithm magnitude spectrum [15]. Once LPC vector is obtained, it is possible to compute Cepstral Coefficients. LPC vector is defined by

$$[a_o a_1 a_2 \ldots a_p]_.$$

LPCC vector is defined by

$$[c_o c_1 c_2 \ldots c_p].$$

LPC vectors are converted to LPCCs using a recursion technique [16]. The recursion is defined with equations (14), (15) and (16):

$$c_0 = ln\sigma^2 \qquad\qquad (14)$$

$$c_m = a_m + \sum_{k=1}^{m-1} \left(\frac{k}{m}\right) c_k a_{m-k} \;\; 1 \leq m \leq p \qquad (15)$$

$$c_m = \sum_{k=1}^{m-1} \left(\frac{k}{m}\right) c_k a_{m-k} \;\; m > p \qquad (16)$$

$\sigma^2$ - gain term in the LPC model

$c_m$ —cepstral coefficients

$a_m$ —predictor coefficients

$k - 1 < k < N - 1$

$p - p$th order

The Cepstral Coefficients, which are the coefficients of the Fourier transform representation of the log magnitude of the spectrum, have been shown to be more robust for audio retrieval than the LPC coefficients [16]. Usually, it is used as a cepstral representation with Q>p coefficients, where Q> (3/2) p.

With enough support of contemporary works [15, 16, 17] reported in allied area of survey, we quantified the 12 dimensional MFCC, LPC and LPCC feature vector, to produce a better query retrieval performance. Further, three parameters were varied to investigate the effectiveness of MFCC, LPC and LPCC in QBSH system. Firstly query excerption is varied from 100% to 60%, and then database size varied from 50 songs to 1000 songs, finally different distance measures were employed. The results of the analysis are elucidated in section 5.

## 4. DISTANCE MEASURES

Euclidean Distance (ED), K-Nearest Neighbour (k-NN) and Dynamic Time Warping (DTW) are used as distance measures. Most important properties of each distance measure are discussed. The ED measure is the standard distance measure between two vectors in feature space with dimension DIM. To calculate the ED measure, sum of the squares of the differences between the individual components of ~x and ~p is computed.

$$d^2_{Euclid}(\vec{x}, \vec{p}) = \sum_{i=0}^{DIM-1} (x_i - p_i)^2 \qquad (17)$$

K-NN is a supervised learning algorithm for matching the query instance based on majority of k-nearest neighbour category. Minimum distance between query instance and each of the training set is calculated to determine the k-NN category. The k-NN prediction of the query instance is determined based on majority voting of the nearest neighbour category. In this approach, for each test audio signal, minimum distance from the test and training audio signal is derived to locate the k-NN category. The ED measure relates the closeness of test and training audio signal. For each test audio signal, the training data set is located with k closest members. From this k-NN, ranks of test audio samples are found.

The DTW is the third measure for finding similarity between two time series which may vary in time. It encourages finding the optimal alignment between two times series in which one time series warped non-linearly by stretching or shrinking it along its time axis. This warping is employed to search corresponding regions or to find out the similarity among the two time series. Figure 5 demonstrates the warping of one times series to another.

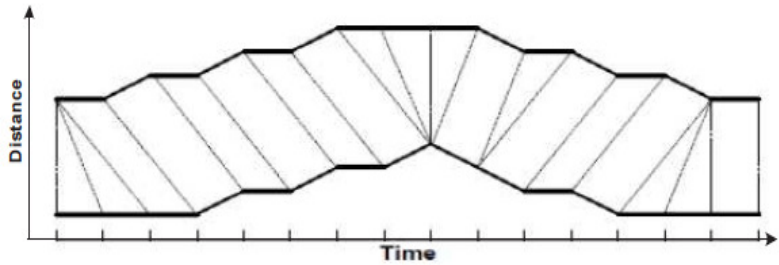

Figure 4: A Warping between two time series.

In figure 5, each vertical line connects a point in one time series to its corresponding similar point in the other time series. The lines have similar values on the y-axis, but have been separated so that vertical lines between them can be viewed more easily. If both of the time series in figure 5 were alike, every line would be straight vertical line as no warping would be necessary to line up the two time series. The warp path distance quantifies the distinction among the two time series after they have been warped together, and the same is measured by the sum of the distances between each pair of points connected by the vertical lines in figure 5. Hence, two time series that are identical apart from localized stretching of the time axis will have DTW distances as zero. The principle of DTW is to compare two dynamic patterns and measure its similarity by calculating a minimum distance between them. The classic DTW is computed as below:

Suppose we have two time series Q and C, of length n and m respectively, where:

$$Q = q_1, q_2, \ldots, q_i, \ldots, q_n$$

$$C = c_1, c_2, \ldots, c_j, \ldots, c_m$$

To align two sequences using DTW, an n-by-m matrix where the (ith, jth) element of the matrix contains the distance d ($q_i$,$c_j$) between the two points $q_i$ and $c_j$ is constructed [17]. Then, the absolute distance between the values of two sequences is calculated using the ED computation:

$$d(q_i, c_j) = (q_i - c_j)^2 \qquad\qquad (18)$$

Each matrix element (i, j) refers to the alignment among the points $q_i$ and $c_j$. Then, accumulated distance is measured by:

$$D(i,j) = min[D(i-1, j-1), D(i-1, j), D(i, j-1)] + d(i, j)$$
$$(19)$$

## 5. RESULTS AND DISCUSSIONS

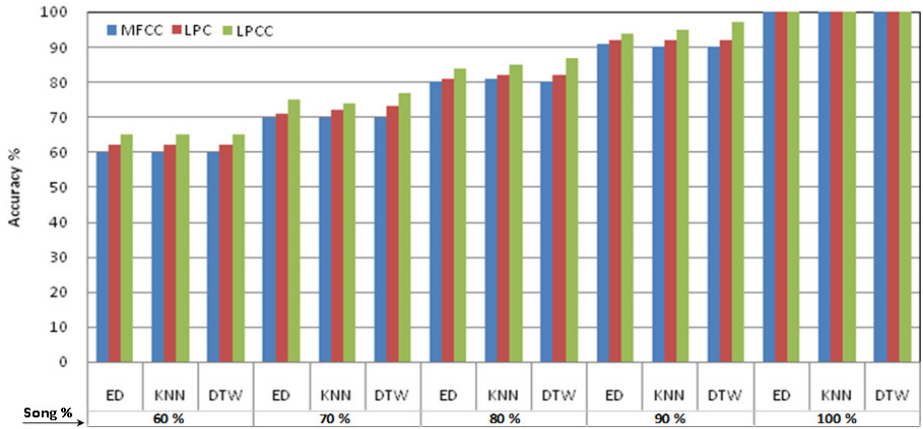

Figure 5: Accuracy % Vs Distance Measures.

In this fragment, the results of similarity measures such as ED, KNN and DTW are discussed under a variety of circumstances such as different query excerption, database size and feature extracting technique. The retrieval accuracy versus different distance measures data is depicted in figure 6. Experiments were conducted based on MFCC, LPC and LPCC features with varying song/hum percentage and different distance measures. It was experimentally determined that song retrieval accuracy increases with increase in song/hum percentage also LPCC out performs slightly because of less complexity and computational time.

One more observation is in a consistent manner; DTW produced slightly better retrieval performance compared to other two distance measures. Because it minimizes the total distance between the respective points of the signal. The retrieval performance varied from 60% to 100% for MFCC, LPC and LPCC features for all three distance measures with query song/hum percentage variation from 60% to 100%.

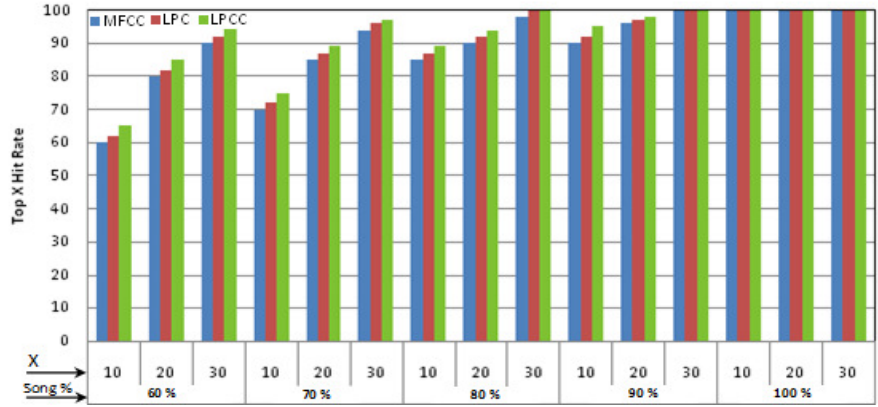

Figure 6: Top X Hit Rate % Vs Song/Hum %.

The impact of X-values on accuracy for MFCC, LPC and LPCC features are portrayed in Figure 7. The top X hit rate is defined as percentage of successful queries and it can be shown mathematically as:

$$Top(X) = \#\{rank(i) : rank(i) \leq X\}/N \qquad (20)$$

where X symbolize top most songs.

The top X hit rate varied from 60% to 100% for MFCC, LPC and LPCC features with X value 10, 20 and 30 respectively. Generally, the results indicate that the LPCC features achieved a little better top X hit rate when compared to MFCC and LPC features for different percentage of query song/hum with three distance measures. From the figure 7, X value 10 was found to be the best, at which system obtained retrieval accuracy in the range 62% to 100% depending on song/hum query excerption specified.

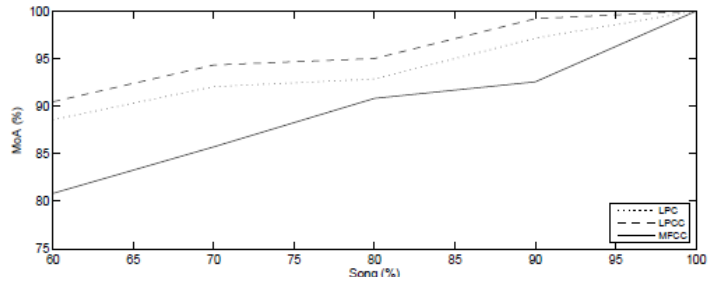

Figure 7: MoA% Vs Query Song/Hum %.

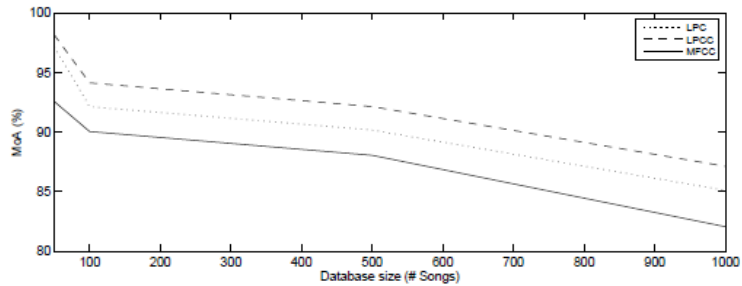

Figure 8: MoA% Vs Database Size (# Songs).

Many different measures for evaluating the performance of QBSH systems have been proposed. The measures require a collection of training and testing samples. For each test scenario and parameter combination the Mean of Accuracy (MoA) is defined as:

$$MoA = \frac{1}{n} \sum_{i=1}^{n} \frac{n - rank(t_i)}{n - 1} \qquad (21)$$

It demonstrates the average rank at which the target was found for each query with a value of 50% describing random, 57% to 67% mediocre, and above 67% good accuracy. We obtained MoA in the range 82 % to 100% with different query excerption. From figure 8, it is found that the MoA increases with increase in query song/hum percentage and LPCC features perform better than MFCC and LPC features. Further figure 9 shows that MoA decreases as database size increases. QBSH analysis exhibits MoA in the range 93% to 98% with different database size.

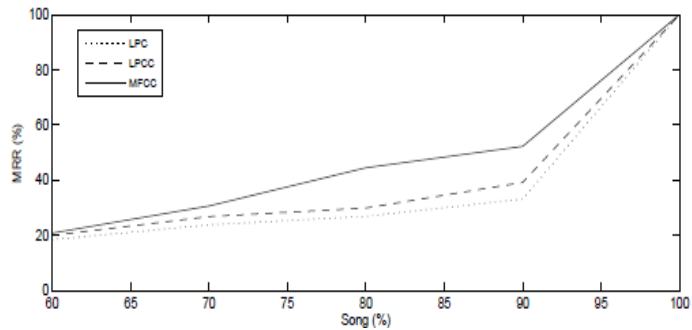

Figure 9: MRR% Vs Query Song/Hum %.

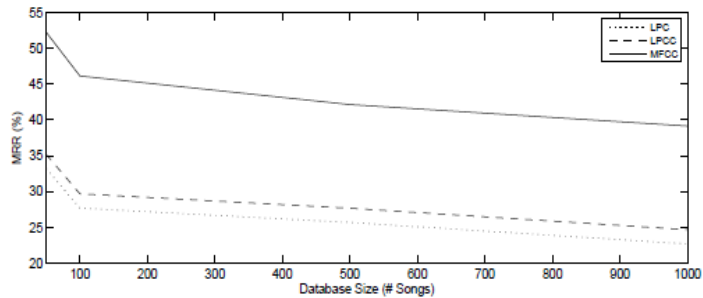

Figure 10: MRR% Vs Database Size (# Songs).

Mean Reciprocal Rank (MRR) is a statistic for assessing any system that produces a list of possible responses to a query, ordered by probability of appropriateness. The reciprocal rank of a query result is the multiplicative inverse of the rank of the first right response. The MRR is stated as the average of the reciprocal ranks of responses for queries Q:

$$MRR = \frac{1}{n} \sum_{i=1}^{n} \frac{1}{rank(t_i)} \qquad (22)$$

The reciprocal value of the MRR refers to the harmonic mean of the ranks. MRR indicates the probability of the target reaching one of the first ranks. The MRR estimate is very much dependent on database size and on 200 song database MRR should be more than 0.2. We obtained MRR in the range 20 % to 100% with different query excerption. From figure 10, it is found that the MRR increases with increase in query song/hum percentage and MFCC features perform better than LPC and LPCC features. Further figure 11 shows that MRR decreases as database size increases. QBSH analysis exhibits MRR in the range 34% to 53% with different database size.

Slag side of inferences observed in the experiment indicates that, the performance is quite comparable with the contemporary [4, 5, 9, 11, 14] works. But, these works employed different music types and databases.

## 6. CONCLUSIONS

This paper attempted to present an approach to QBSH through different query excerption for music search to liberate ranked list. Also different distance metrics are used in the analysis to perform similarity comparisons between melody database and query. The translation of the acoustic input into a symbolic query is crucial for the effectiveness of QBSH system. System works efficiently on MP3 devotional songs based on vocal part and achieved retrieval performance comparable to the state-of-the-art. Proposed analysis exhibits empirical performance by returning the desired song within the top 10 hits 65% of the time and as the top hit 21% of the time on a database with 1000 songs. Our results demonstrate that the size of the training database and song/hum percentage are also main factors that determine the success of the approach. We are able to observe the comprehensiveness of the system for different database sizes in terms of LPCC performing better than other two features. Subsequently there is enough scope for expanding melody database and adapting the system to different singers and songs.

## Authors

Trisiladevi C. Nagavi is currently pursuing Ph D in Computer Science and Engineering under the Visvesvaraya Technical University Belgaum, Karnataka, India. She did her M.Tech in Software Engineering from Sri. Jayachamarajendra College of Engineering Mysore in 2004. She is working as Assistant Professor in Computer Science & Engineering department, Sri Jayachamarajendra College of Engineering, Mysore, Karnataka, India. Her areas of interest are Audio, Speech and Image processing.

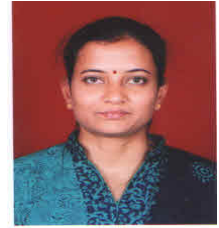

Nagappa. U. Bhajantri has completed Ph D under the University of Mysore, Karnataka, India. He did his M.Tech in Computer Technology (Electrical Engineering Department) from Indian Institute of Delhi, India in 1999. His areas of interest are Image, Video and Melody processing. He is currently working as Professor and HOD of Computer Science and Engineering department in Government Engineering College, Chamarajanagar, Karnataka, India.

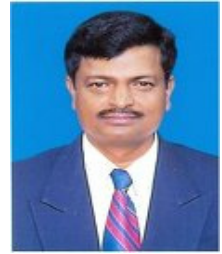